\documentclass{ws-ijmpe}
\usepackage{epsfig}
\newcommand{\oh}{{1\over2}}
\begin{document}

\markboth{Oset et al.}{Chiral unitary dynamics of hadrons ...}

\catchline{}{}{}{}{}

\title{CHIRAL UNITARY DYNAMICS OF HADRONS AND HADRONS IN A NUCLEAR MEDIUM\\
\footnote{chiral unitary dynamics of hadrons and hadrons in a nuclear medium}
}

\author{\footnotesize E. OSET, L. S. GENG, D. GAMERMANN, M.J. VICENTE VACAS, D.
STROTTMAN, K. P. KHEMCHANDANI, A. MARTINEZ TORRES \footnote{oset@ific.uv.es}}

\address{Departamento de F\' isica Te\' orica and IFIC, 
Centro Mixto Universidad de Valencia-CSIC, 
Institutos de Investigaci\'on de Paterna\\
Aptd. 22085, 46071 Valencia, Spain.
\footnote{}\\
}

\author{J. A. OLLER AND L. ROCA}

\address{Departamento de F\'{\i}sica. Universidad de
Murcia.\\
Murcia. E-30071 Murcia.  Spain.\\
}

\author{MAURO NAPSUCIALE}

\address{Instituto de F\'{i}isica, Universidad de Guanajuato\\
Lomas del Bosque 103, Fraccionamiento Lomas del Campestre\\
37150, Le\'{o}n Guanajuato, M\'{e}xico.
}

\maketitle

\begin{history}
\received{(received date)}
\revised{(revised date)}
\end{history}

\begin{abstract}
In this talk I summarize recent findings around the description of axial vector
mesons as dynamically generated states from the interaction of pseudoscalar
mesons and vector mesons, dedicating some attention to the two $K_1(1270)$
states. Then I review the generation of open and hidden charm scalar and axial 
states, and how some recent experiment supports the existence of the new
hidden charm scalar state predicted. I present recent results showing that the 
low lying $1/2^+$
baryon resonances for S=$-1$ can be obtained as bound states or resonances of two mesons
and one baryon in coupled channels. Then show the differences with the 
S=$0$ case, where the $N^*(1710)$ appears also dynamically generated from the
two pion one nucleon system, but the $N^*(1440)$ does not appear, indicating a
more complex structure of the Roper resonance. Finally I shall show how the 
state X(2175), recently discovered at BABAR and BES, appears naturally as a 
resonance of the $\phi K \bar{K}$ system.

\end{abstract}

\section{Introduction}
The combination of nonperturbative unitary techniques in coupled channels with
the QCD information contained in the chiral Lagrangians has allowed one to extend
the application domain of traditional Chiral Perturbation theory to a much
larger range of energies where many low lying meson and baryon resonances
appear. For instance, for the interactions between the members of the lightest 
octet of pseudoscalars, one starts with the chiral Lagrangian of 
ref. \cite{gasser,ulf} and selects the set of channels that couple to certain 
quantum numbers. Then, independently of using either the Bethe-Salpeter 
equation in coupled channels \cite{npa}, the N/D method \cite{nsd} or the 
Inverse Amplitude Method \cite{ramonet}, 
the well known scalar resonances $\sigma(600)$, $f_0(980)$, $a_0(980)$ 
and $\kappa(800)$ appear as poles in the obtained L=0 meson-meson partial 
waves.  These
resonances are not introduced by hand, they appear naturally as a consequence of
the meson interaction and they qualify as ordinary bound states or resonances
in coupled channels. These are states that we call dynamically generated, by
contrast to other states which would rather qualify as $q \bar{q}$ states, such as
the $\rho$ \cite{Pelaez:2003dy}.  Similarly, in the baryon sector, the 
interaction of the pseudoscalar
mesons with baryons of the octet of the proton generates dynamically $1/2^-$
resonances \cite{weise,angels,ollerulf,carmina,jido,ollersolo} and the interaction of the
pseudoscalar mesons with baryons of the decuplet of the $\Delta$ generates $3/2^-$
resonances \cite{lutz,sarkar}. These last two cases can be unified using SU(6)
symmetry as done in \cite{carmenjuan}. This field has proved quite productive
and has been further extended by combining pseudoscalar mesons with vector mesons,
which lead to the dynamical generation of axial vector mesons like the
$a_1(1260)$, $b_1(1235)$, etc \cite{lutzaxial,roca}. Also in the charm sector
one has obtained in this way scalar mesons with charm, like the $D_{s0}(2317)$
\cite{koloscalar,hofmann,guo2,daniel}, axial vector mesons with charm like the
$D_{s1}(2460)$ \cite{lutzaxial,chiangaxial,danielaxial}, or hidden charm scalars
like a predicted $X(3700)$ state and two hidden charm axial states, with
opposite C-parity, one of which corresponds to the $X(3872)$ state. In what
follows we briefly discuss these latter cases. 

Paralelly with these developments investigation has started regarding systems with
three hadrons not studied so far, like two mesons and a baryon, 
or three mesons. A new formalism is developed, based on the Faddeev equations 
and the on shell scattering amplitudes of the different components. This can be
done because one can prove that there are cancellations between the off shell 
part of these
amplitudes and the three body contact forces that originate from the same
chiral lagrangians.  This novel finding should stimulate thoughts around
conventional Faddeev equations which rely upon the full off shell extrapolation
of the amplitudes, and eventually require some three body forces that must 
be put by hand to reproduce the data. I shall report
on several states which are obtained within this new formalism and which can be
associated to well known resonances.

\section{Axial vector mesons dynamically generated}

\begin{figure*}[ht]
\begin{center}
\begin{tabular}{cc}
\includegraphics[scale=0.25]{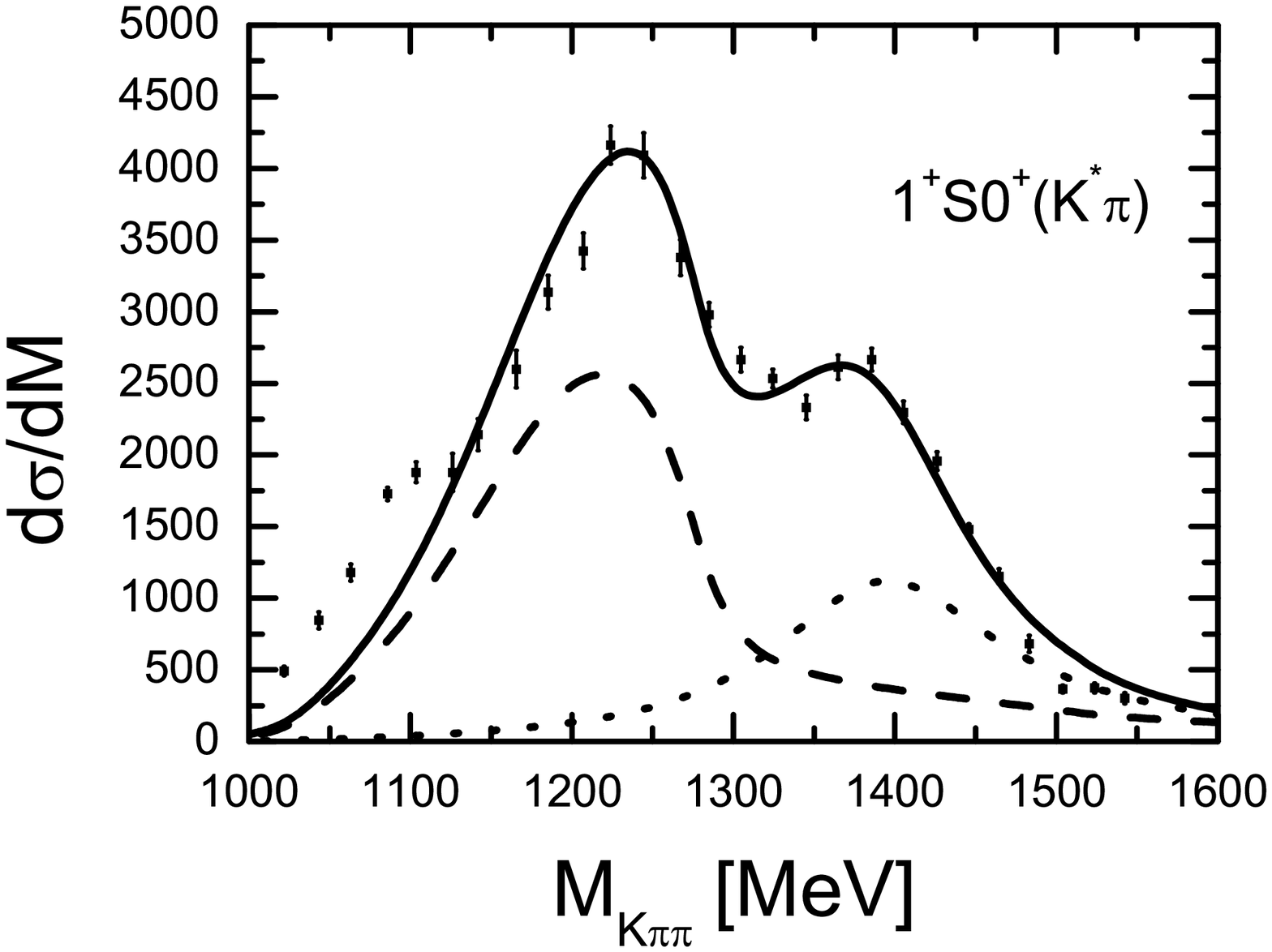} %
&\includegraphics[scale=0.25]{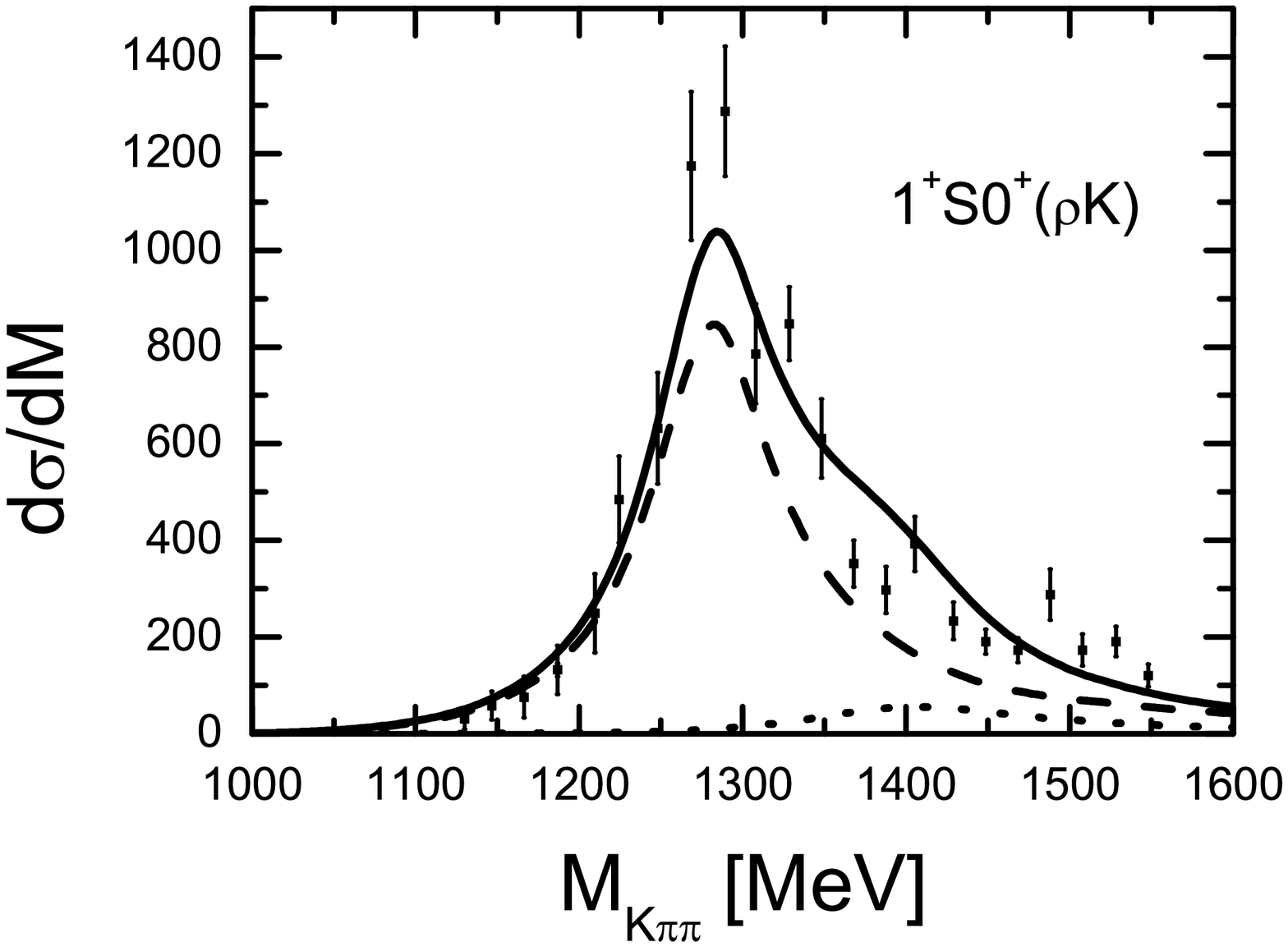}\\
\includegraphics[scale=0.25]{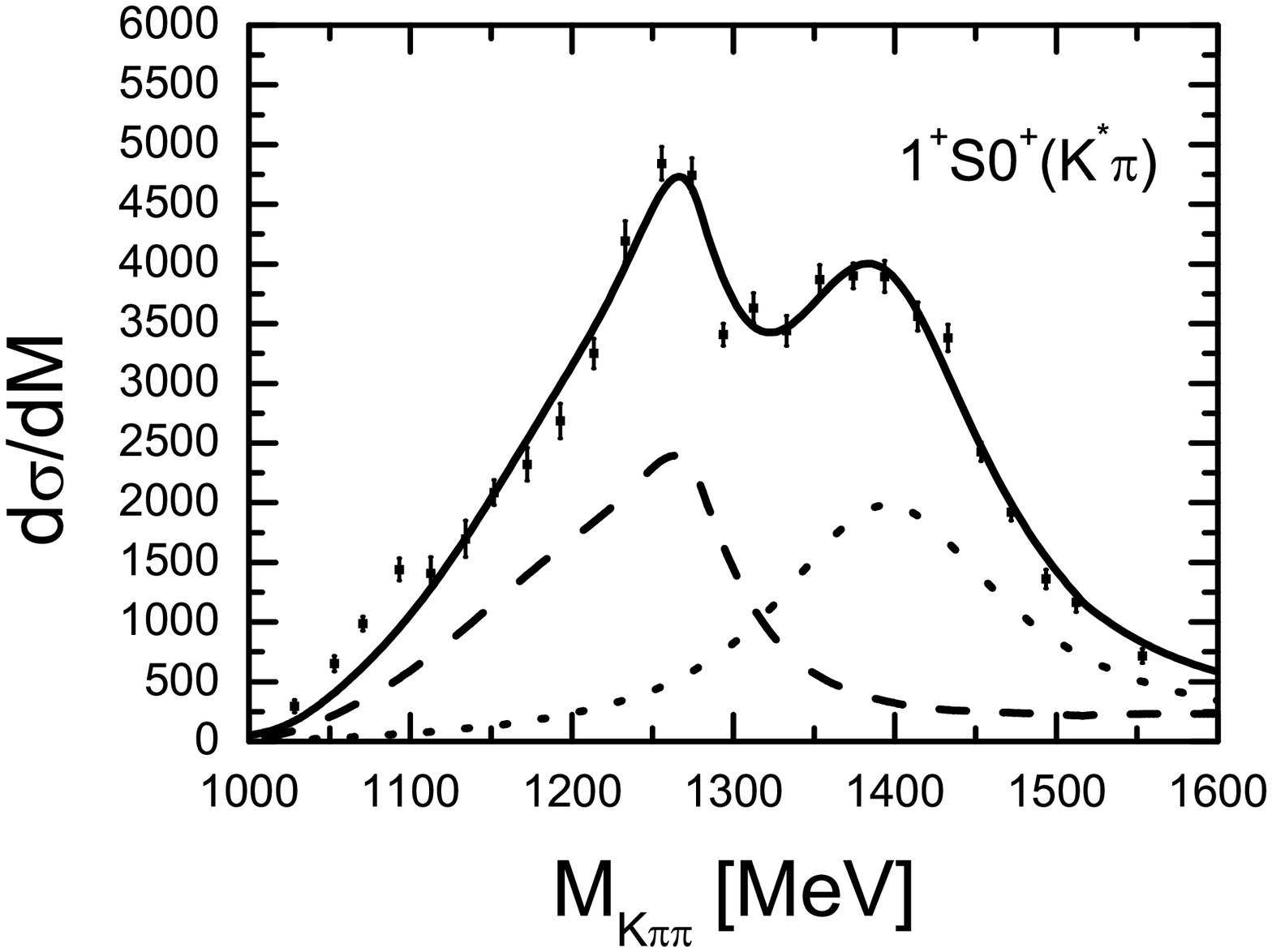}%
&\includegraphics[scale=0.25]{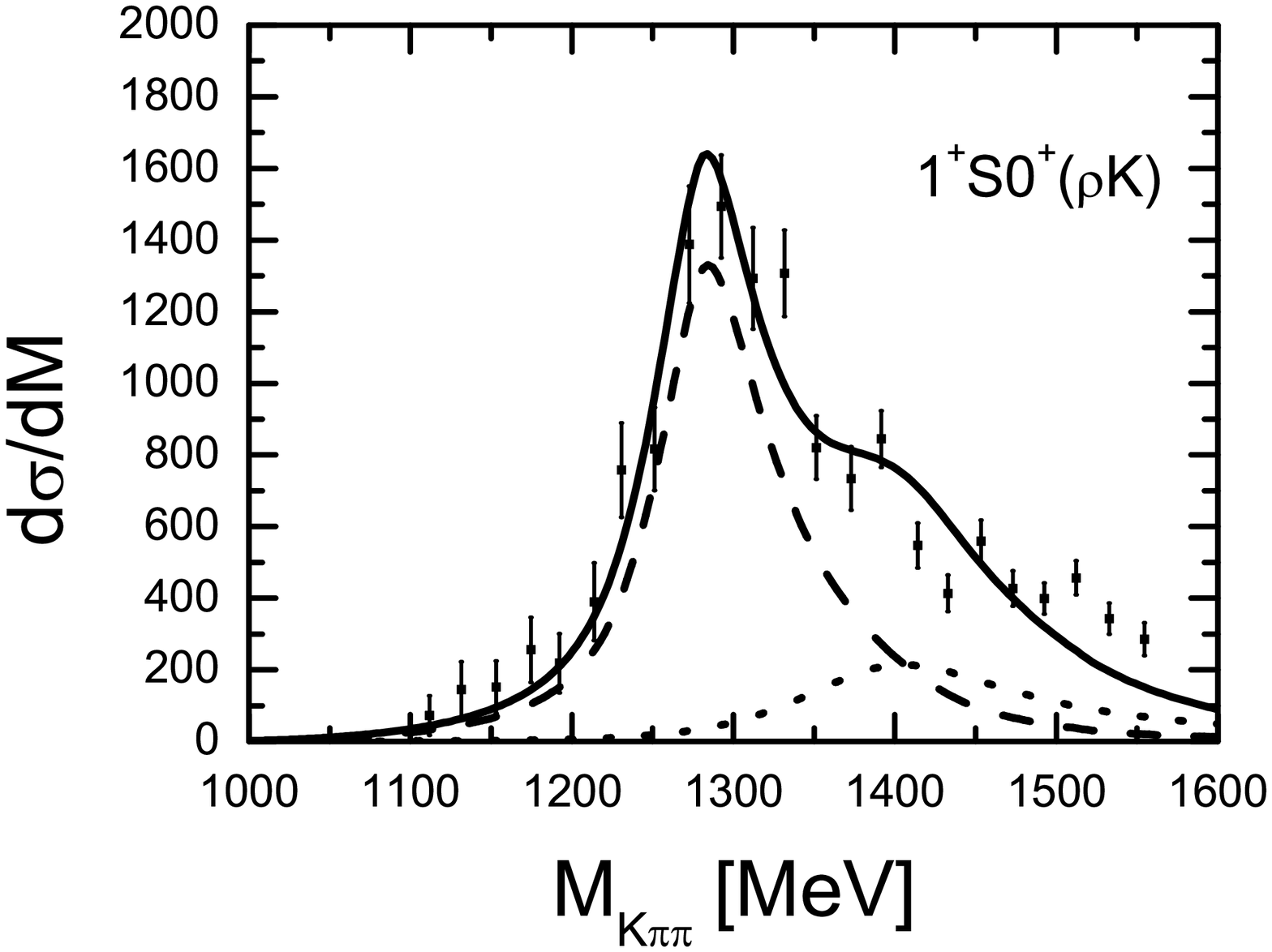}\\
\includegraphics[scale=0.25]{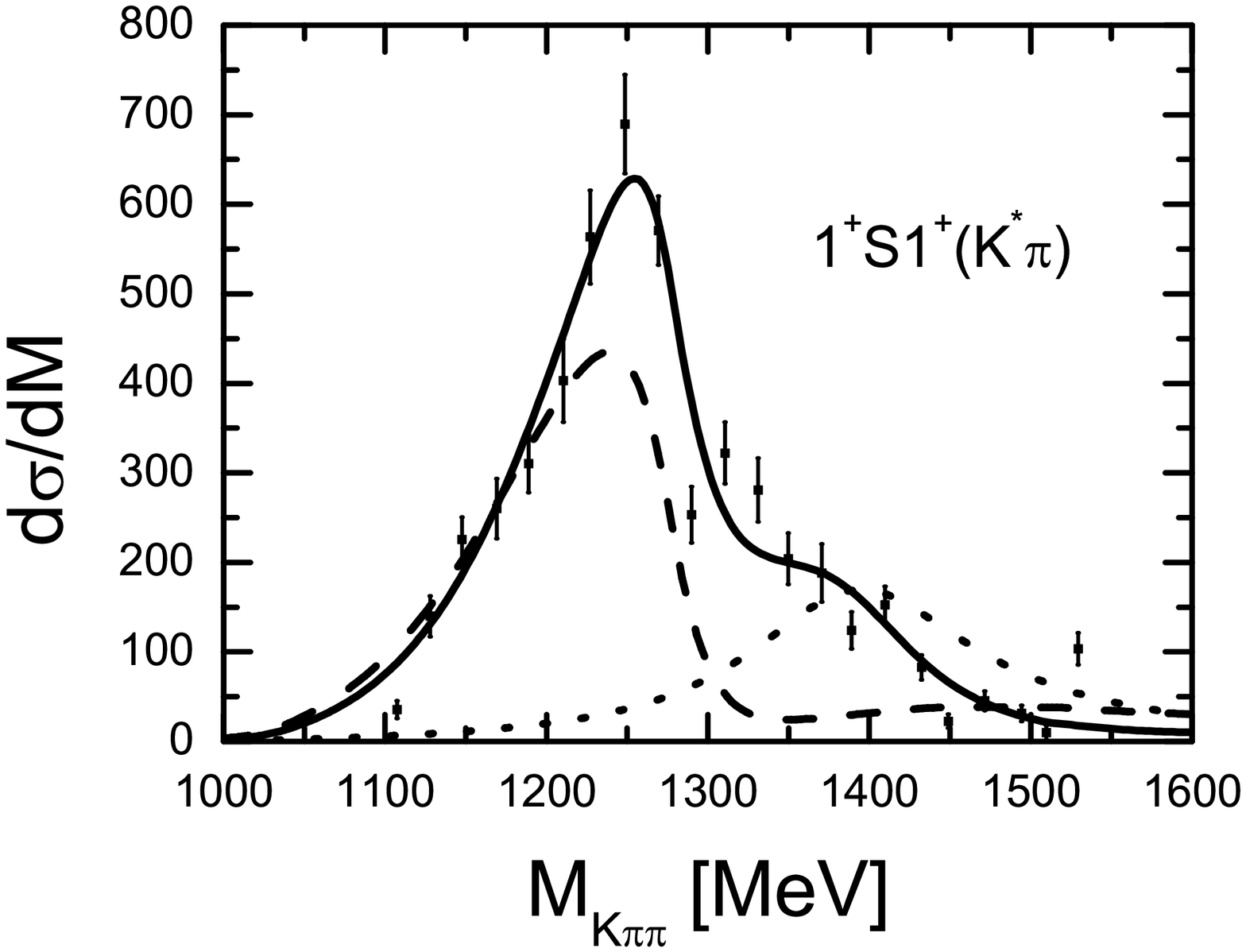}%
&\includegraphics[scale=0.25]{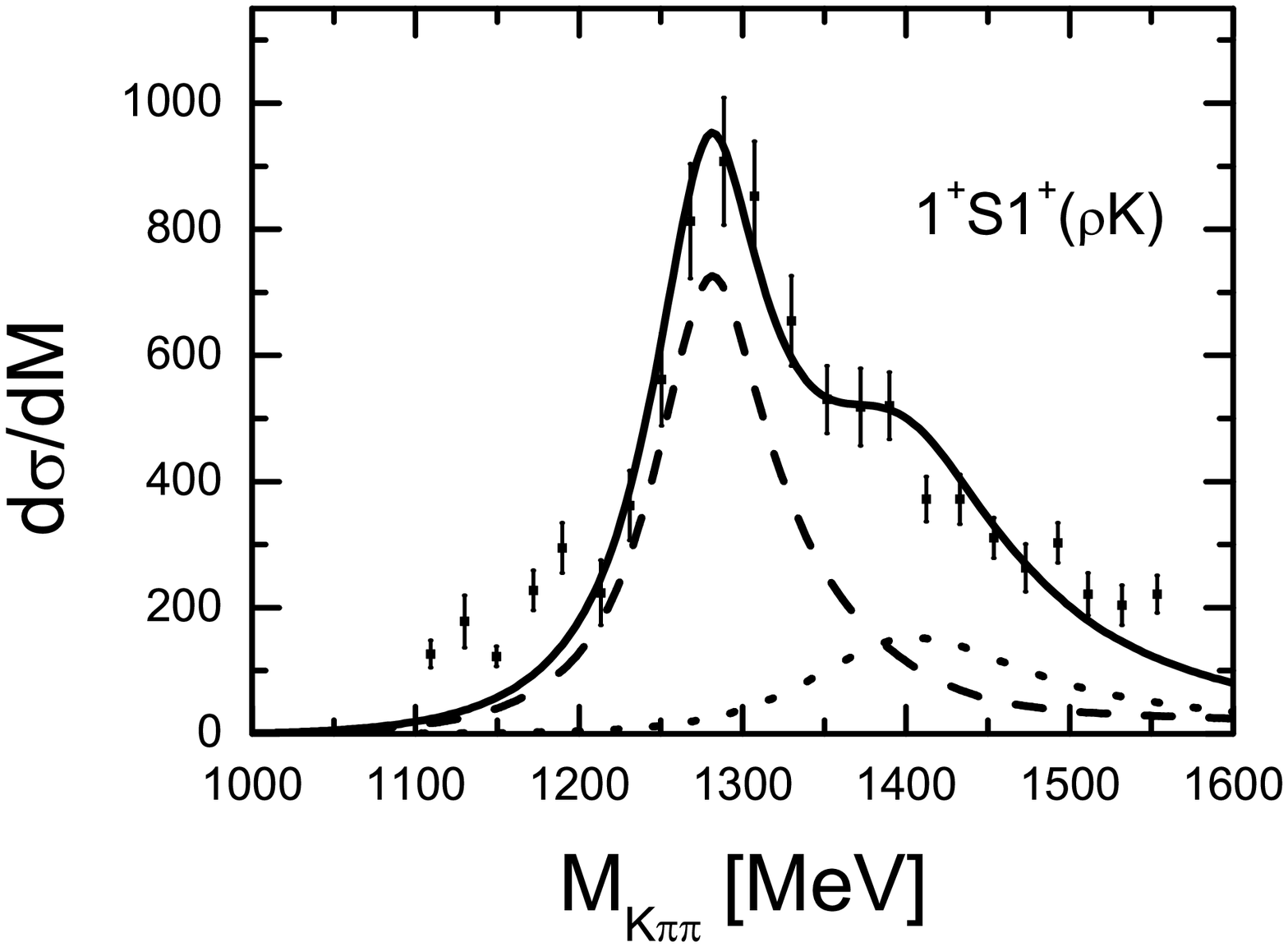}\\
\end{tabular}
\caption{Results for the $ \pi \pi K$ invariant mass distribution in the 
$K^- p \to K^- 	\pi^+  \pi^- p$  reaction. Data in
the upper panels are for $0\le |t'| \le 0.05 ~ GeV^2$ and those in the middle 
and bottom panels for $0.05 \le |t'| \le 0.7 ~ GeV^2$, where $t'$ is the
four momentum transfer squared to the recoiling proton. The data are further 
grouped by $J^PLM^\eta$ followed by the isobar and odd
particle. J is the total angular momentum, P the parity, L the orbital angular 
momentum of the odd particle. $M^\eta$ denotes the magnetic
substate of the $K \pi \pi$ system and the naturality of the exchange.}
\label{fig:CERN3}
\end{center}
\end{figure*}

As shown in detail in \cite{lutzaxial,roca}, starting from  a standard chiral
Lagrangian for the interaction of pseudoscalar mesons of the octet of the $\pi$
and vector mesons of the octet of the $\rho$, and unitarizing in coupled
channels solving the coupled Bethe-Salpeter equations, one obtains the scattering
matrix for pseudoscalar mesons with vectors for different quantum numbers, 
which contains poles that can be associated to known resonances like the 
$a_1(1260)$, $b_1(1235)$, etc. The SU(3) decomposition of $8\times 8$ 
\begin{equation}
8 \times 8 = 1 + 8_s + 8_a + 10 + \overline{10} + 27 
\end{equation}
leads here to two octets, unlike in the case of the interaction of pseudoscalars
among themselves where there is only room for the $8_s$ representation. This is
why here one finds different G-parity states like the $a_1$ and $b_1$,
 the $f_1(1285)$, $h_1(1380)$, plus an
extra $h_1(1170)$ that one can identify with the singlet state. One should then
find two $K_1$ states, which do not have defined G-parity. One might
think that these states are the $K_1(1270)$ and the $K_1(1400)$ states. However
the theory fails to predict a state with such a large mass as the $K_1(1400)$
and with its decay properties. Instead, in \cite{roca} two states were found with 
masses close by,  given, after some fine tuning, by 1197 MeV and 1284 MeV, and 
widths of about 240 MeV and 140 MeV, respectively \cite{geng}.  The interesting 
thing about
these states is that the first one couples most strongly to  $K^* \pi$, while 
the second state couples most strongly to $K \rho$. One could hope that these
two states could be observed experimentally. Indeed, this is the case as was
shown in the recent work \cite{geng} by looking at two reactions which have
either $K^* \pi$ or $K \rho$ in the final state and which clearly show the peak
at different positions, as one can observe in fig. \ref{fig:CERN3}.

 It is interesting to recall that in the experimental analysis done in
 \cite{daum} only one $K_1(1270)$ resonance was included (together with the
 $K_1(1400)$ which shows up at higher energies), but the background was very
 large and the peaks appeared from interference of large background terms rather
 than from the effect of the resonance.   Instead in \cite{geng}, with the
 introduction of the two resonances obtained in our approach and the 
 background generated by the same chiral unitary approach, together with 
  a contribution from
 the $K_1(1400)$ considered phenomenologically, the description of all data in
 fig. \ref{fig:CERN3} follows in a natural way. 

\section{Dynamically generated scalar mesons with open and hidden charm  }

 A generalization to SU(4) of the SU(3) chiral Lagrangian for meson-meson 
 interaction is done in \cite{daniel} to study meson-meson  interaction
 including charm. The breaking of SU(4) is done as in \cite{hofmannbar,mizutani},
 where the crossed exchange of vector mesons is employed as it accounts 
 phenomenologically  for the Weinberg-Tomozawa term in the chiral Lagrangians. 
  With this in mind, when the exchange is due to a heavy vector
 meson the corresponding term is corrected by the ratio of square masses of the
 light vector meson to the heavy one (vectors with a charmed quark).  We also
 use a different pattern of SU(4) symmetry breaking by following the lines  of
 a chiral motivated model with general SU(N) breaking \cite{walliser}.  The picture
 generalizes the model used in \cite{koloscalar,hofmann,guo2}, where only the light vector
 mesons are exchanged. The same states generated in \cite{lutz,guo2} are  also
 generated in \cite{daniel} with some changes, but in addition one obtains
 states with hidden charm. The changes refer to the states of the sextet, which
 in  \cite{daniel} appear rather broad, while these are narrow in other works. In
 table 1 we show the states with charm or hidden charm
 obtained in the present approach. 

\begin{table}
\begin{center}
Table 1: {Pole positions for the model. The column Irrep shows the 
results in the $SU(3)$ limit.} \label{poleposset2}\\
\begin{tabular}{c|c||c|c|c|c|c}
\hline
C& Irrep &S&I$(J^{P})$& RE($\sqrt{s}$) (MeV)& IM($\sqrt{s}$) (MeV)&Resonance ID\\
 & Mass (MeV) & & & & & \\
\hline
\hline
 &          & & & & & \\
1&$\bar 3$&1&0$(0^+)$&2317.25&0&$D_{s0}^*(2317)$ \\
\cline{3-7}
 &2327.96 &0&$\oh(0^+)$&2129.26&-157.00&$D_0^*(2400)$ \\
\cline{2-7}
 & 6   &1&1$(0^+)$&2704.31&-459.50& (?) \\
\cline{3-7}
& 2394.87&0&$\oh(0^+)$&2694.69&-441.89& (?) \\
\cline{3-7}
& -i219.33&-1&0$(0^+)$&2709.39&-445.73&(?) \\
\hline
0&1&0&0$(0^{+})$&3718.93&-0.06&(?)\\
& & & & & & \\
\hline
\end{tabular}
\end{center}
\end{table}
As we can see, the $D_{s0}(2317)$ and $D_0(2400)$ appear in the approach, the
last one at a lower energy than in experiment, but consistent with the data
considering the large width of the state and the theoretical and experimental
uncertainties on the mass. The other three charm states in the table come from a
sextet and they are very broad in our approach ($\Gamma \sim IM(\sqrt{s})$). 

  The very interesting and novel aspect with respect to other theoretical works
 is the heavy state with zero charm. It is a hidden charm state mostly built from
 $D \bar{D}$ and $D_s \bar{D}_s$. The fact that this state has such a narrow
 width in spite of having all the meson-meson states of the light sector open
 for decay, is an interesting consequence of the work, which largely decouples
 the light sector from the heavy one respecting basic OZI rules.  There is no
 experimental information on this state presently, but an
 enhancement of the cross section of the $e^+ e^- \to J/\psi D \bar{D}$ close to
 the $D \bar{D}$ threshold seen in \cite{plakhov} could be interpreted 
  \cite{danielnew} as a consequence of the
 effect of the X(3700), which is a bound state below, but close to, 
 the $D \bar{D}$ threshold. I show this in detail in the next section.
 
 \section{Experimental hints of the X(3700) hidden charm state}
 
\begin{figure}[h] 
\begin{center}
\includegraphics[width=6cm]{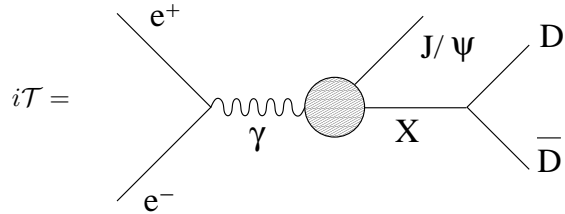}  
\put(-210,40){$i{\cal T}=$} 
\caption{Feynman diagram for the process $e^+e^-\rightarrow J/\psi D\bar
D$} 
\label{diagram} 
\end{center}
\end{figure}

 The reaction $e^+e^-\rightarrow J/\psi D\bar D$ can be described by the
diagram in fig. \ref{diagram} if one assumes that the $D\bar D$ pair comes from a
resonance.

Close to threshold the only part of this amplitude which is strongly energy
dependent is the $X$ propagator and all other parts can be factorized, so that
we can write

\begin{eqnarray}
{\cal T}&=&C {1\over M_{inv}^2(D\bar D)-M_X^2+i\Gamma_X M_X} \label{eq1}
\end{eqnarray}
if we describe the $X$ resonance as a Breit-Wigner type.

The cross section would then be given by an integral over the phase space of
the three particles in the final state \cite{danielnew}.

Assuming that $\cal T$ depends only on the $D\bar D$ invariant mass, one can
evaluate the differential cross section:

\begin{eqnarray} {d\sigma\over dM_{inv}(D\bar D)}&=&{1\over(2\pi)^3}{m_e^2\over
s\sqrt{s}} |\overrightarrow k| |\overrightarrow p| |{\cal T}|^2 \label{dcross}
\end{eqnarray} 
where $s$ is the center of mass energy of the electron positron
pair squared and $|\overrightarrow k|$ and $|\overrightarrow p|$ are given by:

\begin{eqnarray}
|\overrightarrow p|&=&{\lambda^{1/2}(s,M_{J/\psi}^2,M_{inv}^2(D\bar D))\over2\sqrt{s}} \label{pmom} \\
|\overrightarrow k|&=&{\lambda^{1/2}(M_{inv}^2(D\bar D),M_{D}^2,M_{\bar D}^2)\over2M_{inv}(D\bar D)} \label{kmom}
\end{eqnarray}
 Where $\lambda^{1/2}(s,m^2,M^2)$ is the usual K\"allen function.

In the following  we explain how we compare our results to Belle's data.

Belle has measured the differential cross section for $J/\psi D\bar D$, $J/\psi
D\bar D^*$ and $J/\psi D^*\bar D^*$ production from electron positron collision
at center of mass energy $\sqrt{s}$=10.6 GeV \cite{plakhov}. We are going to
study the first case, where the scalar hidden charm state generated in the model
of \cite{daniel} plays a special role. The Belle's measurement produces
invariant mass distributions for the $D\bar D$ that range from
threshold up to 5.0 GeV. Our model is in principle reliable for energies within
few hundreds of MeV from the thresholds, so we compare numerically
our results with the data up to 4.2 GeV.

The experiment measures counts per bin. In the case of a $D\bar D$ pair, the
bins have 50 MeV width, while for the $D\bar D^*$ pair they have 25 MeV. To
compare the shape of our theoretical calculation with the experimental data we
integrate our theoretical curve in bins of the same size as the experiment and
normalize our results so that the total integral of our curve matches the total
number of events measured in the invariant mass range up to 4.2 GeV.
The comparison is made by the standard $\chi^2$ test.

As described in \cite{daniel,danielaxial}, in the heavy sector the model to evaluate
the scattering T-matrix has one free parameter, $\alpha_H$ which is the
subtraction constant in the loop for channels with at least one heavy particle.
In these previous papers this parameter has been fitted so that the pole in the
C=1, S=1, I=0 sector matches the observed state with these quantum numbers
($D_{s0}(2317)$ and $D_{s1}(2460)$ for scalar and axial states, respectively).
The channels in this sector have always one heavy and one light meson, and in
principle one could fit a different $\alpha$ for channels with two heavy
particles. Here we are going to present results for different values of
$\alpha$ in channels with hidden charm (doubly heavy channels). Since we are
working with the C=0 sector, we have also channels involving only light mesons.
These have negligible influence in the pole position of the hidden charm poles
as shown in \cite{daniel}, so we leave $\alpha_L$ constant. The values chosen
for $\alpha_H$ correspond to the natural size \cite{ollerulf}. In terms of an
equivalent cut off to regularize the loop functions, the value $\alpha_H=-1.3$
corresponds to $q_{max}\sim$850 MeV, for two $D$ mesons in a loop. As mentioned
above, the values of $\alpha_H$ taken in the calculation correspond to those
taken in \cite{danielaxial,daniel}. We have taken a range of $\alpha_H$ roughly
around the values $\alpha_H=-1.3$ chosen for the scalar mesons \cite{daniel}.

\begin{figure}[t]
\begin{center}
\includegraphics[width=7.5cm,angle=-0]{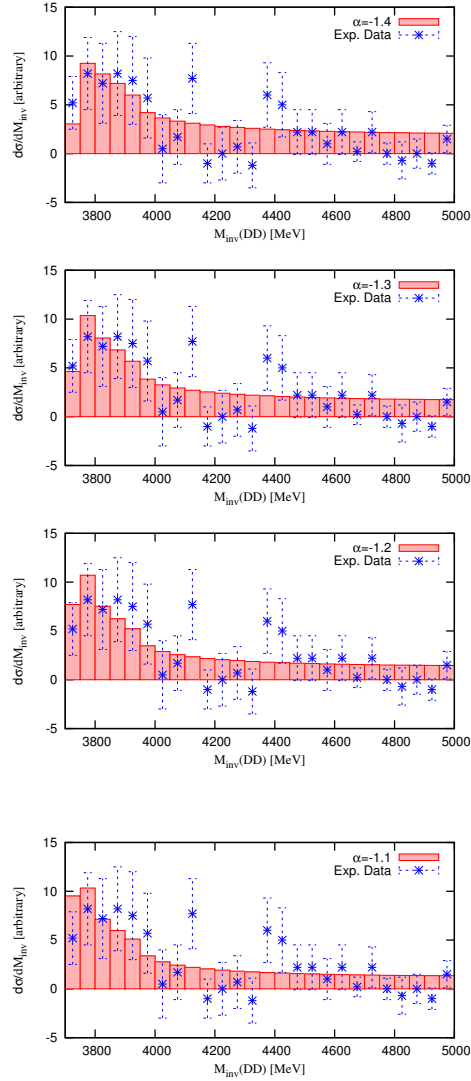} 
\caption{Theoretical histograms compared with data  for $D\bar
D$ invariant mass distribution.} 
\label{fig2}
\end{center}
\end{figure}

In table \ref{tab1} we show results, for different values of $\alpha_H$, of the
pole position of the hidden charm resonance in the scalar sector, and the value
of $\chi^2$ calculated with the data from Belle, with combinatorial background
already subtracted, for all points below 4.2 GeV in the $J/\psi D\bar D$
production. Fig. \ref{fig2} shows plots of our theoretical histograms compared
with experimental data \cite{plakhov}. Note that although we are plotting all
points until 5.0 GeV, only the ones below 4.2 have been used in the calculation
of $\chi^2$ and in the normalization of the theoretical curves.

\begin{table}
\begin{center}
\caption{Results of $M_X$ and $\chi^2$ for different values of $\alpha_H$.} \label{tab1}
\begin{tabular}{c|cc}
\hline
$\alpha_H$ & $M_X$(MeV) & $\chi^2 \over d.o.f$ \\
\hline
\hline
-1.4   & 3701.93-i0.08 & 0.96 \\
-1.3   & 3718.93-i0.06 & 0.85 \\
-1.2  & 3728.12-i0.03 & 0.92 \\
-1.1   & Cusp & 1.11 \\
\hline
\end{tabular}
\end{center}
\end{table}

The $\chi^2$ values obtained in table \ref{tab1} are around 1 or below,
indicating a good fit to the data in all curves. This is in part due to the
large experimental errors, but the clear message is that the presence of a pole
below the $D\bar D$ threshold is enough to reproduce the observed enhancement
of the cross section for this reaction in the $D\bar D$ invariant mass above
threshold. The results of table \ref{tab1} and inspection of fig. \ref{fig2}
show some preference for values of $\alpha_H$=-1.3, -1.2, which would
correspond to the hidden charm scalar with mass slightly above 3700 MeV.

\begin{figure}[h]
\begin{center}
\includegraphics[width=7cm,angle=-0]{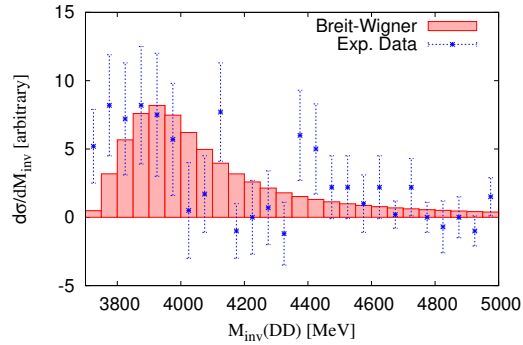} 
\caption{Histograms calculated with Breit-Wigner resonance with mass $M_X$=3880 MeV compared to data.} \label{fig4}
\end{center}
\end{figure}

The peak seen in the experiment has been fitted with a Breit-Wigner like
resonance in \cite{plakhov}, suggesting a new resonance. In order to make the
results obtained here more meaningful, we also perform such a fit and compare
the results. We take the same Breit-Wigner parameters suggested in the
experimental paper, $M_X$=3878 MeV and $\Gamma_X$=347
MeV. We show the
results obtained by fitting a Breit-Wigner form from eq. (\ref{eq1}) in $\cal
T$ of eq. (\ref{dcross}) in fig. \ref{fig4}. Additionally we
calculate $\chi^2$ and find $\chi^2/d.o.f$=2.10 for the $D\bar D$ distribution.
 The value of $\chi^2$
for the $D\bar D$ distribution can be improved if we take different parameters
for the Breit-Wigner resonance. Taking for the fit $M_X$=3750 MeV and
$\Gamma_X$=250 MeV we obtain a value of $\chi^2/d.o.f$=1.12, still slightly
bigger than those obtained in our previous analysis assuming the mechanism of
fig. \ref{diagram} driven by the $X(3700)$ scalar state. 

As a consequence of the discussion, our conclusions would be that for the case
of the broad peak seen in $D\bar D$, the weak case in favor of a new state
around 3880 MeV discussed in \cite{plakhov} is further weakened by the analysis
done here, showing that the results are compatible with the presence of a
scalar hidden charm state with mass around 3700 MeV.

 \section{Dynamically generated axial vector mesons with open and hidden charm}

With the interaction of pseudoscalar mesons with vector mesons in
\cite{danielaxial} one obtains the results shown in Table 2. In addition to
the well known $D_{s1}(2460)$, $D_1(2430)$, $D_{s1}(2536)$ and $D_1(2420)$ (and
all those in the light sector already found in \cite{lutzaxial,roca}) one obtains
new states, which could be observed, although some of them are either too broad
or correspond to cusps. 

\begin{table}
Table 2: Pole positions for the model. The column Irrep shows the results in 
the $SU(3)$ limit. The results in brackets for the $Im\sqrt{s}$ are obtained 
taking into account the finite width of the $\rho$ and $K^*$ mesons. \label{axialsmodel}
\begin{center}
\begin{tabular}{c|c||c|c|c|c|c}
\hline
C& Irrep &S&I$^G(J^{PC})$& RE($\sqrt{s}$) & IM($\sqrt{s}$) &Resonance ID\\
 & Mass (MeV) & & & (MeV)& (MeV)& \\
\hline
\hline
1&$\bar 3$&1&0$(1^+)$&2455.91&0&$D_{s1}(2460)$ \\
\cline{3-7}
 &2432.63 &0&$\oh(1^+)$&2311.24&-115.68&$D_1(2430)$ \\
\cline{2-7}
 & 6   &1&1$(1^+)$&2529.30&-238.56& (?) \\
\cline{3-7}
& 2532.57&0&$\oh(1^+)$&Cusp (2607)&Broad& (?) \\
\cline{3-7}
& -i199.36&-1&0$(1^+)$&Cusp (2503)&Broad&(?) \\
\cline{2-7}
& $\bar 3$ &1&0$(1^+)$&2573.62&-0.07~[-0.07]&$D_{s1}(2536)$ \\
\cline{3-7}
&2535.07&0&$\oh(1^+)$&2526.47&-0.08~[-13]&$D_1(2420)$ \\
& -i0.08& &     &      &     & \\
\cline{2-7}
& 6 & 1&1$(1^+)$& 2756.52&-32.95~[cusp]&(?) \\
\cline{3-7}
&Cusp (2700)&0&$\oh(1^+)$&2750.22&-99.91~[-101]&(?) \\
\cline{3-7}
&Narrow&-1&0$(1^+)$&2756.08&-2.15~[-92]&(?) \\
\hline
0&1&0&0$^+(1^{++})$&3837.57&-0.00&$X(3872)$\\
&3867.59& & & & & \\
\cline{2-7}
&1&0&0$^-(1^{+-})$&3840.69&-1.60&(?)\\
&3864.62& & & & & \\
\hline
\end{tabular}
\end{center}
\end{table}

     Very interesting and novel in the present approach is the generation of the
X(3872) with positive C-parity and another state nearly degenerate
with negative C-parity. It would be interesting to see if a state with negative
C-parity is observed, but the large branching fraction 
\begin{equation}
\frac{B(X \to \pi^+ \pi^- \pi^0 J/\psi)}{B(X \to \pi^+ \pi^-  J/\psi)}=1.0\pm
0.4 \pm 0.3
\end{equation}
indicates either   a very large G-parity (isospin) violation (quite unlikely), or the existence of
another state with different C-parity (G-parity also in this case).

\section{Dynamically generated $1/2^+$ baryon states from the interaction of
two mesons and one baryon}

We discussed before how the low lying $1/2^-$ baryon resonances appear
dynamically generated in the chiral unitary approach. The low lying $1/2^+$
resonances are not less problematic and quark models have difficulties in
reproducing them \cite{glozman}. Experimentally some of them are poorly
understood  and few of them 
 possess four-star status.
Among the rest some resonances are listed with unknown spin parity and two
are controversial in nature. The situation is slightly better with the
$\Lambda$ resonances in the same energy region, except for the $\Lambda$(1600)
and $\Lambda$(1810), where the peak positions and widths, obtained by different
partial wave analysis groups, vary a lot. Many of these S=$-1$ states seem to
have significant branching ratios for three-body, i.e., two meson-one baryon,
decay channels. However, no theoretical attempt has been made to study the
three body structure of these resonances, until recently when a coupled channel
calculation for two meson one baryon system was carried out using chiral
dynamics \cite{epja,MartinezTorres:2007sr}. 

\section{Formalism for the three body systems} 

We take advantage of the fact that there are strong correlations in the meson
baryon sector in L=0, and with S=$-1$ one obtains many $1/2^-$ resonances.
The $\Lambda(1405)\,S_{01}$ ($J^P=1/2^-$) couples strongly
to the $\pi-\Sigma$ and its coupled channels. Considering this we build the
three body coupled channels by adding a pion to combinations of a pseudoscalar
meson of the $0^-$  SU(3) octet and a baryon of the $1/2^+$ octet which couple
to $S=-1$. For the total charge zero of the three body system we get twenty-two
coupled channels. Details can be seen in \cite{epja,MartinezTorres:2007sr}.

To solve the Faddeev equations we write the two body $t$-matrices using unitary
chiral dynamics. These $t$-matrices can be split into an on-shell part,
depending only on the respective center of mass energy,  and an off-shell part,
which is inversely proportional to the propagator of the off-shell particle.
This off-shell part cancels a propagator in the three body scattering diagrams,
leading to a diagram with a topological structure equivalent to that of a three
body force \cite{epja,MartinezTorres:2007sr}. To this, one must add the three 
body forces originating
directly from the chiral Lagrangians. Interestingly, in our case, we find that 
the three body forces from the two sources cancel in the SU(3) limit. 
In a realistic case, we find them
to sum-up to merely 5 \% of the total on-shell contribution of the
$t$-matrices to the Faddeev equations. The formalism is thus developed further
in terms of the on-shell parts of the two body $t$-matrices.

We begin with Faddeev equations 
\begin{equation}
T^i=t^i\delta^3(\vec{k}^{\,\prime}_i-\vec{k}_i)+ t^i g^{ij} T^j + t^i g^{ik} T^k ,
\end{equation}
which, if iterated while neglecting the terms with $\delta^3(\vec{k}^{\,\prime}_i-\vec{k}_i)$,
 corresponding to the disconnected diagrams, will give
\begin{eqnarray}\nonumber 
T^i = t^i g^{ij} t^j + t^i g^{ik} t^k + t^i g^{ij} t^j g^{jk} t^k + t^i g^{ij} t^j g^{ji} t^i
+t^i g^{ik} t^k g^{kj} t^j + t^i g^{ik} t^k g^{ki} t^i + ... .
\end{eqnarray}

In order to convert the Faddeev equations into a set of algebraic equations
 one writes the terms with three
successive interactions explicitly, which already involve a loop evaluation. One
finds technically how to go from the diagrams with two interactions to
those with three interactions and the algorithm found is then used for the next
iterations, leading thus to a set of algebraic equations, which are solved
within twenty two coupled channels.

\begin{figure}[hbt]
\begin{center}
\includegraphics[scale=0.85]{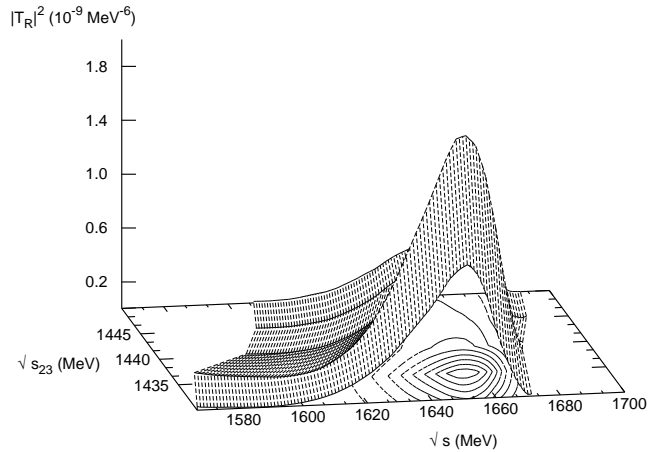}
\caption[]{The $\Sigma$ (1660) resonance in the $\pi^0 \pi^0 \Sigma^0$ channel.}\label{fig1}
\end{center}
\end{figure}

\begin{figure}[t]
\begin{center}
\includegraphics[width=0.85\textwidth]{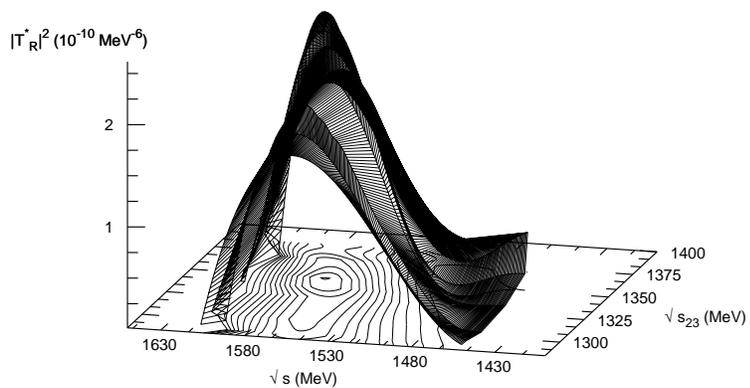}
\caption{\label{fig5}A $\Lambda$ resonance in the $\pi \bar{K} N$ amplitude at 1568 MeV in $I$ = 0, $I_{\pi \bar{K}}$ = 1/2.}
\end{center}
\end{figure}

The resulting Faddeev equations have been solved with the input two body $t$-matrices
obtained by solving the Bethe-Salpeter equation as in 
\cite{npa,angels,inoue}. We find four $\Sigma$ and two $\Lambda$ states as
dynamically generated resonances from a two meson-one baryon system,
which we associate to known resonances of the PDG \cite{pdg}, implying
a strong coupling of the S=$-1$ resonances, in this region, to the three body
decay channels. In Fig. \ref{fig1} we plot the modulus square of the T-matrix as
a function of two variables, the total energy and the invariant mass of
particles two and three.  We show results for one of the resonances, corresponding to the
$\Sigma$(1660) \cite{pdg} found in our study in the squared amplitude for the
$\pi^0 \pi^0 \Sigma^0$ channel, and in fig. \ref{fig5} we show the results for
the $\Lambda(1600)$. In addition to this, we find evidence
for (1) another $1/2^+$ resonance, i.e., the $\Sigma$(1770), (2) for the
controversial $\Sigma$(1620) and (3) for the $\Sigma$(1560), which is listed
with unknown spin-parity \cite{pdg}. In the isospin 0 sector we find evidence
for the $\Lambda$ (1600) and $\Lambda$ (1810).

\begin{figure}
\begin{center}
\includegraphics[width=0.85\textwidth]{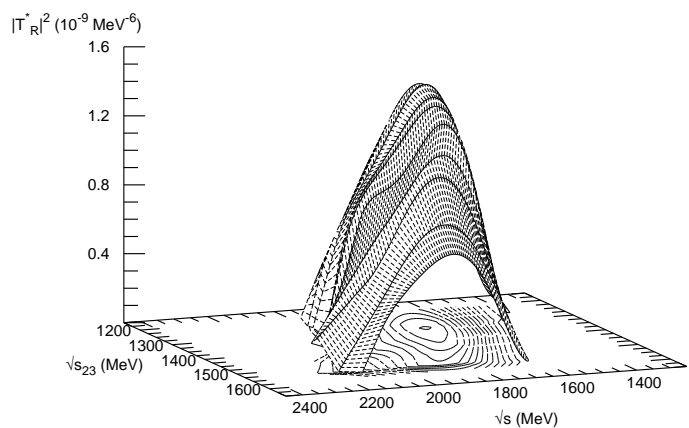}
\caption{\it The squared amplitude for the $\pi \pi N$ system in isospin 1/2 configuration as a function of $\sqrt{s}$ and $\sqrt{s_{23}}$.}
\label{1710_1}
\end{center}
\end{figure}

In a recent paper \cite{Khemchandani:2008rk}
the work has been extended to the sector with $S=0$. There one finds neatly the 
$N^*(1710)$ as a resonant state of $\pi \pi N$, but the Roper $N^*(1440)$ does
not show up, indicating a far more complicated structure, as one may guess from
the study of  \cite{Krehl:1999km}, where the $\pi N$ , $\pi \Delta$  and $\sigma
N$ channels, with interactions additional to those used in the approach 
of \cite{Khemchandani:2008rk} are considered. In this paper one can find in
detail the cancellations between the off shell parts of the amplitudes and the
three body forces coming from the chiral Lagrangians that we have mentioned
above. The signal for the 
$N^*(1710)$ is seen in fig. \ref{1710_1}.

\begin{figure}[ht]
\begin{center}
\includegraphics[scale=0.85]{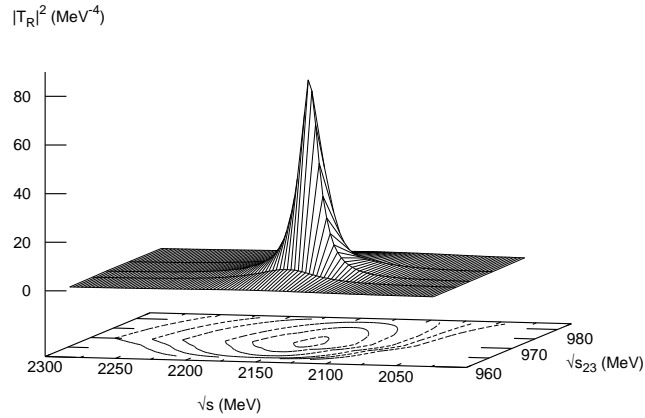}
\caption{The $\phi K \bar{K}$ squared amplitude in the isospin 0 configuration.}
\label{ampsq}
\end{center}
\end{figure}

  Finally we would like to report on the recent work  on the generation
  of the X(2175) resonance found at BABAR in the $e^+ e^- \to \phi f_0(980)$ 
  \cite{Aubert:2006bu,Aubert:2007ur} and also confirmed at BES in the study of the $J/\psi \to \eta \phi f_0(980)$
 decay mode \cite{:2007yt}. The work has been done in \cite{MartinezTorres:2008gy}
 with the same three body scheme using the particles $\phi K \bar{K}$. A neat
 resonance is found around 2150 MeV with a width compatible with experiment, see
 fig. \ref{ampsq}.  As mentioned above, we plot the modulus square of the $T$
 matrix as a function of two variables, the total energy and the invariant mass of
 the $K \bar{K}$ subsystem. We not only find the mass on the right place, but we
 also find the peak of the $K \bar{K}$ invariant mass at the $f_0(980)$ position, indicating that the system of
 three particles is strongly correlated in a $\phi$ and the $f_0(980)$ as found
 in the experiment.  
 
In this latter work we also evaluated the quantitative effect of 
cancellation of the off-shell
part of the amplitudes with the three body forces. If the later are omitted,
keeping the off shell dependence of the amplitudes,
 the peak still appears but
shifted by 40 MeV at lower energies. The cancellation between the off shell 
terms of the amplitudes and the three
body forces has thus a nontrivial consequence if omitted. If we look at it from
a different perspective, we can say that should one had used the full off shell
extrapolation of the amplitudes one would have to add the effects of the three
body force, which in this case would induce a shift of 40 MeV in the mass of
the resonance. 

To conclude, we are finding a
new picture for the low lying $1/2^+$ baryon states with $S=-1$, which largely correspond to
bound states or resonances of two mesons and a baryon. In the $S=0$ sector we
find a clear signal for the $N^*(1710)$, which has a very large branching ratio
for decay into $\pi \pi N$ in the PDG, but not for the $N^*(1440)$. This
negative result for the Roper should rather be interpreted in positive terms as
a clear indication that the $\pi \pi N$ that we study is not the most important
component of the Roper structure, hinting, together with the study of 
\cite{Krehl:1999km}, at a very complex structure for the $N^*(1440)$. In the
three meson sector a clear peak could be seen for the three body interacting
system $\phi K \bar{K}$, strongly correlated around the $\phi f_0(980)$,
reflecting
the strong coupling of the $f_0(980)$ resonance to $K \bar{K}$, the main
building block of the $f_0(980)$ as a dynamically generated resonance.

\section*{Acknowledgments}

L. S. Geng wishes to acknowledge support from the Ministerio de Educacion in the
program of Doctores y Tecnologos extranjeros and D. Strottman in the one of
sabbatical. A. Martinez and D. Gamermann from the Ministry of
Education and Science and K. Khemchandani from the HADRONTH project for the EU.
 This work is partly supported by
DGICYT Contract No. BFM2003-00856, FPA2007-62777  and the
E.U. FLAVIAnet network Contract No. HPRN-CT-2002-00311. This
research is part of the EU Integrated Infrastructure Initiative
Hadron Physics Project under Contract No. RII3-CT-2004-506078. The work of
 M. N. was supported by CONACyT M\'{e}xico under project
50471-F.

\end{document}